\documentclass[12pt]{article}
\newcommand{\ba}{\begin{array}}
\newcommand{\ea}{\end{array}}

\newcommand{\be}{\begin{equation}}
\newcommand{\ee}{\end{equation}}
\newcommand{\bea}{\begin{eqnarray}}
\newcommand{\eea}{\end{eqnarray}}
\newcommand{\beal}{\setcounter{letter}{1} \begin{eqnarray}}
\newcommand{\eeal}{\addtocounter{equation}{1} \end{eqnarray}}

\newcommand{\larrow}{\,\,\,\,\hbox to 30pt{\rightarrowfill}
\,\,\,\,}
\newcommand{\slarrow}{\,\,\,\hbox to 20pt{\rightarrowfill}
\,\,\,}

\newcommand{\bm}{\bibitem}

\begin{document}

\begin{titlepage}
\renewcommand{\thefootnote}{\fnsymbol{footnote}}
\renewcommand{\baselinestretch}{1.3}
\medskip

\begin{center}
{\large {\bf A Partially Gauged Fixed Hamiltonian for Scalar Field Collapse }}
\\ \medskip  {}
\medskip

\renewcommand{\baselinestretch}{1}
{\bf
R. G. Daghigh $\sharp$
J. Gegenberg $\dagger$
G. Kunstatter $\sharp$
\\}
\vspace*{0.50cm}
{\sl
$\dagger$ Dept. of Mathematics and Statistics and Department of Physics,
University of New Brunswick\\
Fredericton, New Brunswick, Canada  E3B 5A3\\
}
{\sl
$\sharp$ Dept. of Physics and Winnipeg Institute of
Theoretical Physics, University of Winnipeg\\
Winnipeg, Manitoba, Canada R3B 2E9\\
}


\renewcommand{\baselinestretch}{1}

{\bf Abstract}
\end{center}
We derive a partially gauge fixed Hamiltonian for black hole formation via real scalar field collapse. The class of models considered includes many theories of physical interest, including  spherically symmetric black holes in $D$ spacetime dimensions. The boundary and gauge fixing conditions are chosen to be consistent with generalized Painleve-Gullstrand coordinates, in which the metric is regular across the black hole future horizon. The resulting Hamiltonian is remarkably simple and we argue that it provides a good starting point for studying the quantum dynamics of black hole formation.

{\small}
\vfill
\hfill  
\today
\end{titlepage}
\section{Introduction}

It is widely believed that black hole thermodynamics provides important clues about the structure of the underlying quantum theory of gravity. Spherically symmetric black holes provide a rich testing grounds for analyzing these clues. Moreover, recent work \cite{carlip,solodukhin,carlip2}  suggests that despite the absence of gravitational dynamics, the spherically symmetric sector may contain sufficient structure to describe the origin of black hole entropy.

Perhaps the most significant feature of black hole thermodynamics is its apparent universality. The famous Bekenstein formula for black hole entropy seems to be valid for a very large class of black holes, irrespective of the underlying classical or microscopic quantum dynamics. It is therefore important to understand the classical and quantum structure of black holes generically, and to find new approaches to analyze them. 

The purpose of this letter is to present the results of a novel Hamiltonian analysis that may be very useful in the analysis of spherically symmetric black hole formation via scalar field collapse. The general formalism we use is based on  2-D dilaton gravity but the class of models we consider describes many specific black holes of physical interest, including the spherically symmetric sector of higher dimensional Einstein gravity.

In order to have a system  of equations that is valid at, or even below, any event horizons that might form, we look to Painleve-Gullstrand (PG) coordinates. These were used several years ago by Kraus and Wilczek \cite{kw} to derive a self-interaction correction to black hole radiation in the collapse of spherical shells. Subsequently, Louko {\it et al}\cite{louko_PG} analysed in detail the Hamiltonian dynamics of spherical shell and dust collapse using PG boundary conditions. More recently PG coordinates were used in Hamiltonian analysis of spherically symmetric collapse in 4-d Einstein gravity\cite{hw3}, and to examine the quantization of generic single horizon black holes\cite{gks,kl}.

A crucial ingredient in our analysis is that we do not completely fix the gauge. As in \cite{gks,kl}, only the spatial diffeomorphisms are fixed, so that one is left with a midi-superspace model containing a single Hamiltonian constraint that generates time diffeomorphisms. The proposed analysis of the quantum dynamics will therefore require imposing this Hamiltonian constraint on physical states and then solving for the time evolution generated by the remaining Hamiltonian. The latter is non-vanishing even on the constraint surface because of the presence of the usual boundary terms needed to make the classical variational principle well defined. Although completely reduced models are in principle simpler due to the absence of an infinity of unphysical modes, the resulting Hamiltonians tend to be very complicated, highly non-linear and non-local\cite{ssg_old}. Moreover, since the degrees of freedom that account for black hole entropy may reside in diffeomorphisms that become physical in the presence of black hole boundary conditions \cite{carlip} one may be ``throwing out the baby with the bath water'' by completely gauge fixing prior to quantizing. 

The most striking result of our derivation is the  simplicity of the final, partially reduced Hamiltonian. It consists of three terms. The first is purely gravitational, being proportional to the mass observable. The second term is essentially the matter energy density, which contains only scalar field variables. Finally there is a single term that couples the gravitational mode to the scalar degrees of freedom. The coupling is a fourth order polynomial in the phase space variables, and only second order in the momenta. We will argue that this Hamiltonian system provides an excellent starting point for studying the quantum dynamics of black hole formation.

\section{Classical Action and Solutions}

The formalism we will use to describe generic single horizon black holes is that of 2-D dilaton gravity (for a review see \cite{2dreview}). This approach has been successfully used to study spherically symmetric scalar field critical collapse in any dimension\cite{critical collapse}. The action is: 
\be
S_G[g,\phi]=\frac{1}{2G}\int dx dt \sqrt{-g}\left(\phi R(g)+{V(\phi)\over l^2}\right).\label{dgaction}
\ee
where $l$ is a parameter with dimensions of length which is generally taken to be the Planck length in the theory. The matter action is:
\be
S_M[\psi,g,\phi]=-{1\over 2}\int d^2x\sqrt{-g}h(\phi)|\nabla\psi|^2
\label{matter action}
\ee
Different models are distinguished by two functions of the dilaton field; the dilaton potential $V(\phi)$ and the coupling $h(\phi)$ between the matter field and the dilaton. This class of models describes a large number of physically relevant theories, including the spherically symmetric sector of higher dimensional Einstein gravity. In the case of four dimensions, for example, $\phi\sim r^2$ is the area of the invariant 2-sphere at radius $r$, $V\sim 1/\sqrt{\phi}$ and $h\sim \phi$. For details of the Hamiltonian analysis of the vacuum sector, see for example\cite{exact}. 

 The vacuum theory has no propagating modes and is exactly solvable. There exists a one parameter family of static solutions, which can be represented in Schwarzschild like coordinates $\tilde{x}=l \phi$ as:
\be
ds^2= -(j(\phi) -2GMl)dt_s^2+(j(\phi)-2GMl)^{-1}d\tilde{x}^2
\label{metric solution}
\ee
where $j(\phi) \equiv \int^{\phi}_0 d\phi V(\phi)$ and $M$ is the mass. 

This metric is not asymptotically flat, but, for spherically symmetric gravity in $d=n+2$ spacetime dimensions, it is related to the $(r,t)$ components of the physical, asymptotically flat higher dimensional metric $ds^2_{phy}$ by the conformal reparametrization:
\be
ds^2={j(\phi)}ds_{phys}^2=j(\phi)\left(-(1-{2GMl\over j(\phi)})dt_s^2 + 
            (1-{2GMl\over j(\phi)})^{-1}dr^2\right)
\label{physical metric}
\ee
where the radial coordinate is related to the dilaton by:
\be
dr = l{d\phi\over j(\phi)}\propto \phi^{1/n}
\ee
Note that this conformal reparametrization corresponds to a local field redefinition which converts the action in Eq. (\ref{dgaction}) to its more standard form of spherically symmetric gravity involving the physical metric and containing a kinetic term for the dilaton field. This field redefinition is singular at $\phi=0$, but this surface must in any case be excluded from the manifold in order to represent black hole spacetimes. In fact $\phi=0$ will be treated as a boundary at which suitable boundary conditions must be imposed.

We assume that $j(\phi)$ is monotonic, so that there is at most one horizon, and without loss of generality that it vanishes at $\phi=0$. These conditions guarantee the existence of an apparent horizon in the classical vacuum solution.

It will be useful to consider a coordinate system in which the physical metric is in PG form, i.e.
\be
ds^2= j(\phi)\left(-dt^2 + (dx+\sqrt{2GMl\over j(\phi)}dt)^2 \right)
\label{P-G metric}
\ee
where the time coordinate is related to the Schwarzschild time by:
\be
dt= dt_s + {\sqrt{2GMl\over j(\phi)}}{d\tilde{x}\over j(\phi)- 2GMl}
\ee

In this coordinate system the physical (asymptotically flat) spatial metric is constant and the entire metric is regular at the horizon. The slicing can therefore be extended from spatial infinity all the way to the singularity at $\phi=0$. It is therefore well suited for the study of horizon formation and evolution.

\section{Hamiltonian Analysis}

An analysis of the vacuum theory has been presented in \cite{exact}. Here we present the canonical analysis of the full theory (i.e. with matter), using the metric as parametrized in modified ADM form:
\be
ds^2=e^{2\rho}\left(-\sigma^2 dt^2+\left(dx+N dt\right)^2\right).\label{metric}
\ee
We assume that the 2-D spacetime manifold $M_2=R\times\Sigma$, where $\Sigma$ is a spacelike 
slice.  Note that the exponential parameterization for the conformal mode is traditionally used because it manifestly restricts the conformal mode to be positive definite.

As usual the lapse and shift functions $\sigma$ and $N$, respectively, are lagrange multipliers that give rise to the Hamiltonian and diffeomorphism constraints in the theory. The momenta of the remaining fields are given by the following:
\bea
\dot{\rho}&=& N\rho'+N' - \sigma G\Pi_\phi \label{momenta-rho}\\
\dot{\phi}&=&N\phi'-\sigma G \Pi_\rho  \label{momenta-phi}\\ 
\dot{\psi}&=&N\psi' + {\sigma \Pi_\psi\over h(\phi)}
\label{momenta-psi}
\eea
where the prime denotes partial differentiation with respect to the spatial coordinate $x$.
The Hamiltonian is:
\be
H=\int dx \left(\sigma{\cal G} + {N}
 {\cal F}\right) + H_{B}
\label{hamiltonian1}
\ee
where ${\cal G}$ is the Hamiltonian constraint,
\be
{\cal G} = {\phi''\over G} -{\phi'\rho' \over G} - G \Pi_\phi\Pi_\rho -{e^{2\rho}\over2G}{ V(\phi)\over l^2} + {\cal G}_M
\label{ham1}
\ee
with
\be
{\cal G}_M = {1\over 2} \left( {\Pi_\psi^2\over h(\phi)} +
   h(\phi)(\psi')^2\right)
\label{cal G}
\ee
and $\cal F$ is the diffeomorphism constraint:
\be
{\cal F} = \rho' \Pi_\rho - \Pi_\rho' + \phi' \Pi_\phi
     +\psi' \Pi_\psi
\label{diffeo1}
\ee
$H_B$ is a boundary term required to make the variational principle well defined. It will be determined by the boundary conditions on the fields, which will be discussed in the next subsection.

For future reference we note that the linear combination:
\bea
\tilde{\cal G} &:=& {l e^{-2\rho}\phi' {\cal G}} + lG e^{-2\rho}\Pi_\rho {\cal F}\nonumber\\
 &=& -{\cal M}' + l {e^{-2\rho}\phi' \over 2}\left(
  {\Pi_\psi^2\over h(\phi)} +
   h(\phi)(\psi')^2
  +{2G\Pi_\rho\Pi_\psi\psi'\over \phi'}
\right) 
\label{newham}
\eea
yields a new hamiltonian constraint in terms of the mass function:
\be
{\cal M} = {l\over 2G}\left(e^{-2\rho}((G\Pi_\rho)^2 -(\phi')^2)
  +{j(\phi)\over l^2}\right)
\label{mass function}
\ee
${\cal M}$ is called the mass function because it approaches a constant at spatial infinity, where it is equal to the ADM mass of the solution. In the presence of matter, $\tilde{\cal G}$ can be integrated to yield a simple and physical constraint that relates the mass function at any spatial coordinate $\phi$ to the integrated matter stress energy:
\be
{\cal M} = M_0 + \int^x_0 l {e^{-2\rho}\phi' \over 2}\left(
 \left( {\Pi_\psi^2\over h(\phi)} +
   h(\phi)(\psi')^2\right)
  +{G\Pi_\rho\Pi_\psi\psi'\over \phi'}
\right) dx 
\label{integrated mass}
\ee

With appropriate boundary conditions, to be discussed in the next section, and corresponding Hamiltonian boundary term needed to make the variational principle well defined, the equations of motion for the momenta are:
\bea
G\dot{\Pi}_\phi&=& - \sigma'' - (\sigma\rho')' +{\sigma\over2 l^2} e^{2\rho} {dV\over d\phi} +(NG\Pi_\phi)' \\
G\dot{\Pi}_\rho&=&(GN\Pi_\rho)' -(\sigma\phi')' +\sigma e^{2\rho} {V(\phi)\over l^2}\\
\dot{\Pi}_\psi &=& (\sigma h(\phi) \psi')' +(N\Pi_\psi)'
\eea

\section{Partial Gauge Fixing and Boundary Conditions}
As in \cite{hw3,gks,kl} we choose a partial gauge and boundary conditions that ensure PG metric at infinity and in vacuum. In \cite{hw3} a timelike gauge fixing condition was imposed, and the Hamiltonian constraint was set strongly to zero. We choose instead to fix the spatial coordinate by adding the constraint:
\be
\chi = \phi' - j(\phi) \sim 0
\label{chi}
\ee
As usual, this is imposed weakly (after the Poisson brackets are calculated). In order for this to be a good gauge fixing condition, it must be second class with at least one of the two original constraints. In fact it is second class with the diffeomorphism constraint:
\be
\left\{\chi(\lambda),{\cal F}(\tau)\right\} =
    \int dx \tau j(\phi)\left(\lambda' + {dj(\phi)\over d\phi}\lambda\right)
\ee
which is invertible for all field configurations on the space of test functions $\lambda$ that vanish on both boundaries.

Consistency requires that $\chi$ must be preserved in time,
\be
\dot{\chi} = \left\{\chi,H\right\}=0
\ee
which gives us the condition on the lagrange multipliers:
\be
\sigma G\Pi_\rho =N\phi'
\label{dot chi}
\ee

Since all the phase space variables besides $\Pi_\phi$ commute with $\chi$ the Dirac brackets for $\rho, \Pi_\rho, \psi, \Pi_\psi$ are the same as the Poisson brackets. This means that we can set $\chi$ and ${\cal F}$ strongly to zero in the Hamiltonian, and obtain the correct equations of motion using the ordinary Poisson brackets. 

After some algebra one can show that the reduced Lagrangian is:
\be
{\cal L}= \int dx \left (\Pi_\rho \dot{\rho} +\Pi_\psi \dot{\psi}\right) -\int dx {\sigma e^{2\rho}\over  j(\phi)}\tilde{\cal G} -\int dx ({\sigma e^{2\rho}\over j(\phi)}{\cal M})'
\label{partially reduced L}
\ee
where $\tilde{\cal G}$ is defined in (\ref{newham}). We have added the boundary term needed for slicings that  approach PG coordinates at both boundaries. The partially reduced equations of motion are:
\bea
\dot{\rho}&=& lG \left( {\sigma e^{2\rho} \over j(\phi)} \right)' e^{-2\rho} \Pi_\rho + lG \sigma {\Pi_\psi \psi'\over j(\phi)}\\
G \dot{\Pi}_\rho&=& l \left( {\sigma e^{2\rho} \over j(\phi)} \right)'\left(G^2 e^{-2\rho} \Pi_\rho^2-e^{-2\rho} j^2/l^2\right) \nonumber\\
    & & +G \sigma \left({\Pi^2_\psi\over h} + h(\psi')^2
           +2{G\Pi_\rho\Pi_\psi\psi'\over \phi'}\right)\\
\dot{\psi}   &=& {\sigma lG \Pi_\rho\psi'\over j(\phi)} + { \sigma \Pi_\psi\over h(\phi)}\label{reduced psi2}\\
\dot{\Pi}_\psi &=& (\sigma h(\phi)\psi')' +\left({\sigma l G\Pi_\rho\Pi_\psi\over j(\phi)}\right)'
\eea

In order to discuss boundary conditions we need to specify the functions $j(\phi)$ and $h(\phi)$ that essentially define the theory. In order to admit single horizon black hole solutions, $j(\phi)$ must be monotonic and without loss of generality we assume it vanishes at $\phi=0$, which therefore corresponds to the singularity. For simplicity we assume $j(\phi)=\phi^a$ where $0<a<2$. It was shown in \cite{cadoni} that this latter condition is necessary for the theory to describe black holes. We also assume that the coupling $h(\phi)$ takes the simple form $h(\phi)=\phi^b$ where $b>0$. For spherically symmetric gravity in $D=2+n$ dimensions, $a=1-1/n$ and $b=1$. 
Under these circumstances suitable set of boundary  and fall-off conditions are given as follows. At spatial infinity, we require:
\bea
e^{2\rho}&\to& j(\phi)\left(1+ O( \phi^{-\gamma_\infty})\right)\nonumber\\
G\Pi_\rho &\to& \sqrt{2GM_\infty j/l}\left(1+ O(\phi^{-(\gamma_\infty-a)})\right)\nonumber\\
{\sigma}&\to& \sigma_\infty\left(1+ O( \phi^{-(\gamma_\infty+a/2-1)})\right) \nonumber\\
\psi&\to& O(\phi^{-b/2-\epsilon_\infty})\nonumber\\
\Pi_\psi&\to& O(\phi^{b/2-\epsilon_\infty})
\eea
where $\gamma_\infty>1-a/2$, and again $M_\infty$ and ${\sigma}_\infty$ are parameters. 


At the origin, $\phi = 0$, the fall off conditions require a bit more care. 
It turns out that the slicings can be extended to the origin in the presence of a singularity by requiring:
\bea
e^{2\rho}&\to& j(\phi) \left(1+ O( \phi^{1-a/2+\gamma_0})\right)\nonumber\\
G\Pi_\rho &\to& \sqrt{2GM_0 j/l}+ O( \phi^{1+\gamma_0})\nonumber\\
\sigma &\to& O( \phi^{2-a+\gamma_0})\nonumber\\
  \psi&\to& O(\phi^{1/2-a-b/2 +\epsilon})\nonumber\\
\Pi_\psi &\to& O(\phi^{-1/2+b/2+\epsilon})
\eea
where $\gamma_0$ and $\epsilon$ can take on any positive value. In the above, $M_0$ is the mass of the singularity. Note that since $\sigma(0)=0$, we are restricting the spatial slices to approach the same PG slice at the origin. This requires $M_0$ to be independent of time, since the spatial slice does not evolve there. The slicings can, however, evolve arbitrarily at the horizon, so that these boundary conditions are suited to the study of horizon formation and evolution prior to singularity formation. 

It can be verified that these boundary conditions guarantee that there are no additional contributions to the surface terms in the action, the Hamiltonian and Liouville terms in the action are finite, and that they are preserved by the dynamical evolution equations. In fact, it is the consistency of the latter that forces us to restrict to spatial slices that do not evolve at $\phi=0$.

Ideally one would like to be able to find more general fall-off conditions that allow
a description of not only horizon formation, but also singularity formation. That is, one should start with a spacetime and scalar field which are initially regular at $r=0$. At late times, some time after horizon formation, one expects a spacelike singularity at $r=0$. In principle, our formalism should allow a description of such a
transition by fixing the behavior of the constant of integration
$M_0(t)$ in Eq.(\ref{integrated mass}), which is in principle an
arbitrary function of time.  In our choice of parameterization, a non-zero value for $M_0(t)$ signals the presence of the curvature singularity at the origin. At large times, as the scalar field falls to the origin, $M_0(t)$ should approach the ADM mass. Unfortunately, we have been unable to find consistent boundary conditions that allow non-zero, time varying $M_0(t)$.  This is currently under investigation.


\section{A Canonical Transformation}

The simplicity of the partially reduced Hamiltonian is particularly striking after the following canonical phase space variables for the remaining gravitational field:
\bea
X(x):= e^\rho\nonumber\\
P(x):= e^{-\rho}G\Pi_\rho
\label{can trans}
\eea
The Poisson brackets now read:
\be
\{X(x),P(y)\}=G\delta(x,y)
\ee
The full Hamiltonian in this case takes a very simple and suggestive form:
\bea
H(X,P,\psi,\Pi_\psi)&=&\int dx\left(-{\sigma X^2 \over 2 j(\phi)} {\cal M}' +\sigma {\cal G}_M +\sigma l  \frac{XP \psi'\Pi_\psi}{ j(\phi)}  \right)+\int dx ({\sigma X^2\over j(\phi)}{\cal M})'\nonumber\\
\label{ham2}
\eea
where 
\be
{\cal M}= {l\over 2G}\left(P^2 - {(\phi')^2\over X^2}+{j(\phi)\over l^2}\right)
\ee
and $\cal{G}_M$ is defined in (\ref{cal G}).

Eq.(\ref{ham2}) is the main result of the present paper.  It has some remarkable features. First of all, the Hamiltonian has decomposed into three terms: the first term involves only the gravitational phase space variables, the second only the matter variables, while the third term provides the coupling between them.  Moreover, the coupling term is very simple in that it is quartic in the dynamical variables. Note that this in part is a consequence of our partial gauge fixing: the function $\phi(x)$ is no longer dynamical but a fixed function of the spatial coordinates.

The key operators that appear in the Hamiltonian are:
$K:=X^2/4$, $R:=P^2-{(\phi')^2\over X^2}$,$D:=XP/2$, $(\psi')^2$, $(\Pi_\psi)^2$,  and 
$\Pi_\psi \psi'$. Note that $K,R$ and $D$ obey an $SO(2,1)$ algebra, so that the  
operators $D$ and $K$ give rise to classically conserved charges under time evolution generated by $R$. This algebra is known to be anomalous at the quantum level and has been extensively studied using a variety of techniques\cite{camblong,esteve}.

\section{Conclusions: Toward the Quantum Dynamics of Black Hole Formation}

Equation (\ref{ham2}) provides an intriguing starting point for the study of the quantum dynamics of gravitational collapse, where one might follow a procedure similar to the one outlined in Husain and Winkler\cite{hw3}:
\begin{enumerate}
\item Find a suitable quantum representation for $\hat{H}$.
\item Fix the lapse function thereby implicitly choosing a time coordinate.
\item Find a suitable initial state $\left|\Psi(t_0)\right>$ that solves the constraint equation and
describes the quantum version of some (semiclassical) initial scalar field configuration.
\item Evolve the initial state (numerically if the representation allows) via infinitesimal time steps via the equation:
\be
\left|\Psi(t_0+\Delta t)\right> = (1 + i \hat{H}\Delta t)\left|\Psi(t_0)\right>
\ee
\item At each time step check whether the expectation value of the future null expansion operator
vanishes somewhere on the spatial slice. That is one looks for a value of $x=x_{hor}$ for which:
\be
\left<\psi(t)|\theta_+(x_{hor})|\psi(t)\right>=0
\ee
where 
$\theta_+(x)$ is a suitably quantized version of the classical future null expansion\cite{hw2,gks}:
\be
\theta_+(x)= P - {\phi'\over X}
\ee
One would expect that in different gauges $x_{hor}$ would take on different values, but that
the presence, or not, of a horizon would be invariant.
\end{enumerate}

The key step in the procedure for quantum evolution is to start with a suitable 
representation which allows the constraint to be solved and the Hamiltonian 
evolution to be evaluated with finite computational resources. In \cite{gks} 
the Hamiltonian constraint in the vacuum case was solved in the Bohr 
representation in terms of a set of eigenstates $\left.|M\right>$ of the operator 
$\hat{{\cal M}}$. The Bohr quantization scheme arises naturally in the context of loop quantum gravity (see \cite{shadow} for a review) but has also been recently advocated as useful for singularity resolution in the quantization of purely geometrical variables\cite{hw1}. The states found in \cite{gks} had interesting properties, including the 
fact that they induced a specific (mass dependent) discretization of space 
exterior to the horizon. This spectrum approached the continuum  far from 
the horizon as well as for macroscopic black holes near the horizon. It 
was argued that they provide a reasonable basis for the Hilbert space of 
physical states of an isolated black hole. A basis of mass eigenstates was 
also obtained in \cite{kl} using a Schr$\ddot{\rm{o}}$dinger representation. In this 
representation as well interesting quantum properties were observed near 
the horizon. 

Irrespective of which representation is used,  Eq.(\ref{ham2}) in conjunction with the results of \cite{gks,kl} suggests a natural basis for studying
the quantum dynamics of black hole formation. The first step would be to 
find a complete set of eigenstates, which we will denote generically by 
$\left|k\right>$, of the operator ${\cal G}_M$ which is the Hamiltonian density for 
the matter fields.  Given the structure of Eq.(\ref{ham2}) one could then start 
with the following representation for the time dependent quantum states.
\be
\left|\Psi(t)\right>= \sum_{M,k} C_{M;k}(t)\left|M;k\right>
\ee
where $\left|M,k\right>:=\left|M\right>\otimes\left|k\right>$ is a direct product of eigenstates of the two operators. Clearly the action of the first two terms in (\ref{ham2}) is relatively simple to compute, and all the interesting dynamics will be generated by the interaction term, which is also relatively simple as a quantum operator.

Although this procedure is highly non-trivial, it does on the surface appear to be tractable. We are currently working on implementation via the Bohr representation.

\section{Acknowledgments}
This work was supported in part by the Natural Sciences and Engineering Research Council of Canada. The authors are extremely grateful to Viqar Husain for guidance and many helpful discussions during the initial stages of this work. GK thanks Jorma Louko for useful conversations and the University of Nottingham for its hospitality during the completion of part of this work.

\end{document}